\documentclass[12pt]{article}
\usepackage{graphicx}
 \oddsidemargin  = - 7 mm 
 \evensidemargin = - 7 mm
\input{epsf.tex}
\newcommand{\rts}{ \sqrt s}
\begin{document}
\date{}
\title{$\eta$ in the nuclear medium within a chiral unitary approach} 
\author{
 {\large T.~Inoue\thanks{e-mail: inoue@ific.uv.es}~ and E.~Oset} \\
 {\small Departamento de F\'{\i}sica Te\'orica and IFIC, 
  Centro Mixto Universidad de Valencia-CSIC }\\
 {\small Institutos de Investigaci\'on de Paterna,
  Apdo. correos 22085, 46071, Valencia, Spain }
       } 
\maketitle
\begin{abstract}
 The self-energy of an $\eta$ meson in the nuclear medium
 is calculated in a chiral unitary approach. 
 A coupled channel Bethe-Salpeter equation is solved 
 to obtain the effective $\eta$-$N$ interaction in the medium.
 The base model reproduces well the free space
 $\pi$-$N$ elastic and inelastic scattering 
 at the $\eta$-$N$ threshold or $N^*(1535)$ region. 
 The Pauli blocking on the nucleons, binding potentials for the baryons
 and self-energies of the mesons are incorporated,
 including the $\eta$ self-energy in a self-consistent way.
 Our calculation predicts about $-54 -i29$ MeV for
 the optical potential at normal nuclear matter
 for an $\eta$ at threshold but also shows a strong energy dependence
 of the potential.
\end{abstract}

\section{Introduction}

 The nuclear medium effects on meson properties are interesting 
 and have been investigated extensively.
 The data from relativistic heavy ion collision at CERN 
 \cite{daniref1,daniref2,daniref3} and at lower energies
 at Bevalac \cite{daniref4,daniref5}, may indicate either
 a lowering of the $\rho$ meson mass or a large broadening of its width.
 In the near future, experiments
 at GSI(HADES collaboration \cite{daniref6,daniref7}) could
 clarify the situation by providing better statistics and mass resolution.
 The recent discovery of the deeply bound state
 of $\pi^-$ in Pb \cite{zakiref1,zakiref2}
 reveals an upward shift of the pion mass of about 20 MeV, and confirms
 the repulsive S-wave interaction of the $\pi^-$ in nuclear matter
 already established from studies of the bulk pionic atom
 data \cite{ericson,garcia,batty}.
 The low energy magnitudes of the $\eta$-$N$ interaction
 and the properties of the $\eta$ meson in the medium, 
 are still open questions.
 The $\eta$ mesic nuclei are expected to provide such informations
 and are searched at several facilities.
 For example, the 
 $^7\mbox{Li}(\mbox{d},^3\mbox{He})^6_{\eta}\mbox{He}$,  
 $^{12}\mbox{C}(\mbox{d},^3\mbox{He})^{11}_{\eta}\mbox{B}$  and 
 $^{27}\mbox{Al}(\mbox{d},^3\mbox{He})^{26}_{\eta}\mbox{Mg}$ 
 reactions are investigated at GSI \cite{gillitzer}.

 In this paper, we study the properties of the $\eta$ in the medium
 from a theoretical point of view.
 We evaluate the effective S-wave $\eta$-$N$ interaction in the medium 
 in a chiral unitary approach.
 The self-energy of $\eta$ is obtained by summing
 the $\eta N$ interaction in the medium 
 over the nucleons in the Fermi sea.
 The present work follows closely the method used in \cite{ramos}
 to evaluate the $\bar K$ nucleus optical potential 
 in a self-consistent way.	
 In the next section, we explain our approach.
 We show our input related with the nuclear medium in section 3.
 The results are given in section 4.
 Section 5 is devoted to the conclusions.

\section{Chiral unitary approach to the $\eta$ self-energy}

\begin{figure}[t]
  \centerline{ 
   \epsfysize = 35 mm  \epsfbox{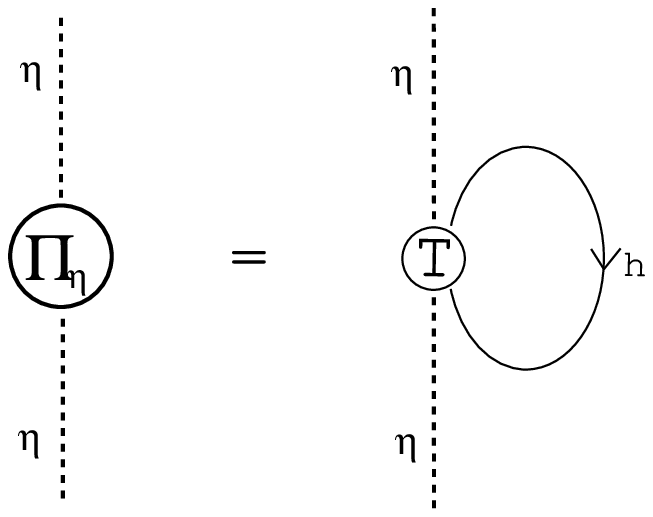}
              }
 \caption{Diagrammatic representation of the $\eta$ self-energy from
   S-wave interaction with nucleons.}
 \label{fig:etasw}
\end{figure}

 We want to obtain the self-energy of the $\eta$ meson 
 in nuclear matter at various densities $\rho$,
 as a function of the $\eta$ energy $k^0$ and the momentum $\vec k$
 in the nuclear matter frame.
 In this paper, we calculate it by means of 
\begin{equation}
     \Pi_{\eta}(k^0, \vec k~; \rho) 
     =   
     2 \! \int^{k_F} \! \!  \frac{d^3 \vec p_n }{(2\pi)^3}~ 
     T_{\eta n}(P^0, \vec P ~; \rho) \times 2
 \label{eqn:selfint}
\end{equation}
 where $\vec p_n$ and $k_F$ are the momentum of the neutron 
 and the Fermi momentum at density $\rho$ respectively, 
 and  $T_{\eta n}(P^0, \vec P ~; \rho)$ is 
 the $\eta$-neutron in-medium S-wave interaction,
 with the total 4-momentum of the system $(P^0,\vec P)$ in the 
 nuclear matter frame, 
 namely $P^0=k^0+E_n(\vec p_n)$ and $\vec P=\vec k + \vec p_n$.
 Here, the isospin symmetry, $T_{\eta p} = T_{\eta n}$, is assumed and
 the amplitude is summed over nucleons in the Fermi sea
 as shown in Fig.\ref{fig:etasw}.
 We shall be concerned about the S-wave $\eta$ self-energy.
 At low energies of the $\eta$ this part of the potential
 is largely dominant.

 We evaluate the in-medium amplitude $T_{\eta n}$ 
 in a chiral unitary approach.
 For this purpose we follow the model for the free space $\pi$-$N$ and
 coupled channels scattering of ref.\cite{inoue},
 which reproduces the experimental data of $\pi N$ scattering
 up to energies above the $N^*(1535)$ region. 
 A similar chiral approach which covers a wider energy range
 in isospin 1/2, although with more free parameters, 
 is also done in ref.\cite{nieves}.

 In ref.\cite{inoue} the Bethe-Salpeter equation is considered 
 with eight coupled channels including two $\pi\pi N$ states, namely
 \{$\pi^- p$, $\pi^0 n$, $\eta n$, $K^0 \Lambda$,
   $K^+ \Sigma^-$, $K^0 \Sigma^0$,
   $\pi^0 \pi^- p$, $\pi^+ \pi^- n$\}.
 The kernels for the meson-baryon two-body sector
 are taken from the lowest order chiral Lagrangians
 and improved by applying a form factor corresponding to a vector meson
 exchange in the t-channel.
 The kernels for $\pi N \leftrightarrow \pi\pi N$ transitions 
 are determined so that they account for 
 both the $\pi N$ elastic and $\pi N \to \pi\pi N$ processes.
 That model reproduces well the the $\pi N$ scattering amplitudes,
 especially in isospin 1/2, for the center of mass energy energies
 from threshold to 1600 MeV.  
 It reproduce also the $\pi^- p \to \eta n$ cross section
 at the region where the P-wave contribution is negligible.
 In this coupled channels approach,  
 the model also provides the $\eta$-$n$ interaction in free space
 and generates dynamically the $N^*(1535)$ resonance providing
 the width and branching ratios for its decay in good agreement with
 experiment, among them the $\eta N$ branching ratio which is
 quite large for that resonance.
 The agreement of the model with the different available data
 around $N^*(1535)$ resonance region and the adequate description 
 of the properties of the resonance, in particular the strong coupling
 to the $\eta N$ state, give us confidence that the model is rather
 accurate to make predictions on the $\eta N \to \eta N$ interaction.
 Therefore we use this model as a base to
 take nuclear matter effects into account.

\begin{figure}[t]
 \centerline{ 
 \epsfysize = 30 mm  \epsfbox{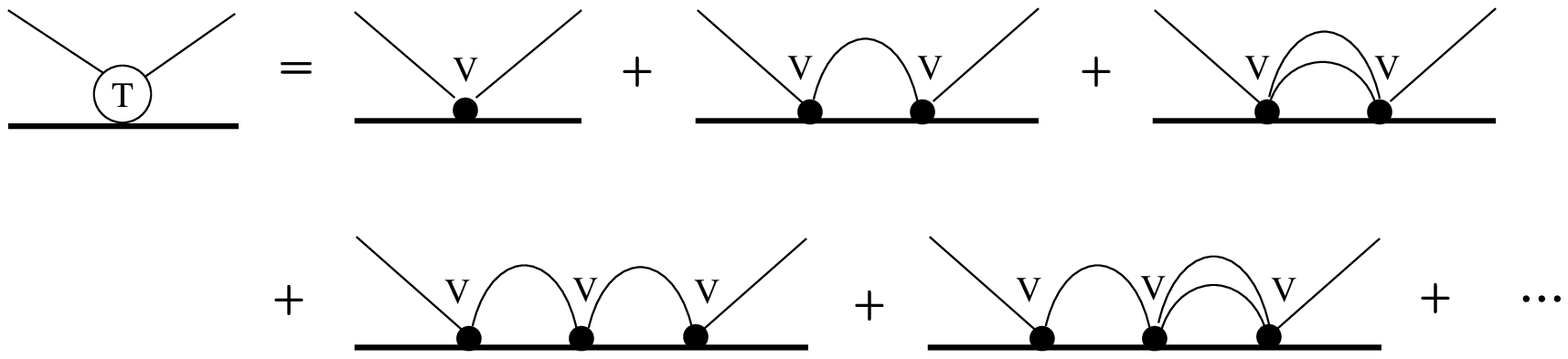}
              }
 \caption{Diagrammatic representation of the Bethe-Salpeter equation.}
 \label{fig:bseq}
\end{figure}

 Analogously to the free space case,  $T_{\eta n}(P^0, \vec P ~; \rho)$ is
 given by one matrix element of the matrix equation
 in the space of  coupled channels
\begin{equation}
  T(P^0, \vec P ~; \rho) =  
  \left[ 1 - V(\rts) G(P^0, \vec P ~; \rho) \right]^{-1}  V(\rts)
\label{eqn:bseq}
\end{equation}
 which is the solution of Bethe-Salpeter equation(Fig.\ref{fig:bseq}),
 where the kernel matrix $V$ have been factorized on shell
 as shown in ref.\cite{angels} and ref.\cite{oller}, 
 and $G$ is a diagonal matrix of loop functions.
 The kernel has nothing to do with nuclear matter and does not depend
 on density and hence can be parameterized in terms of 
 only the invariant energy $\rts$ or $s \equiv (P^0)^2 - \vec P^2$. 
 One can find the explicit form of it in ref.\cite{inoue}. 
 The effects of the nuclear medium are taken into account
 through the loop functions $G_l(P^0, \vec P ~; \rho)$, 
 which describe the propagation of intermediate states in the medium.

 The meson baryon loop functions for the free space
 are given by the integral
\begin{equation}
   G_l(P^0, \vec P)
   =
   \int \! \! \frac{d^4 q}{(2 \pi)^4} 
   \frac{M_l}{E_l(\vec P-\vec q)} 
   \frac{1}{P^0 - q^0 - E_l(\vec P-\vec q)+ i \epsilon} 
   \frac{1}{ (q^0)^2 - {\vec q\,}^2 - m_l^2 + i \epsilon}
 \label{eqn:labint} 
\end{equation}
 which is the zero density limit of the loop function in medium. 
 The generalization to the case where the initial 
 meson baryon system is not in the CM frame ($\vec P \ne \vec 0$),
 is a necessity in order to evaluate the meson baryon amplitude for
 arbitrary meson and baryon momenta in the frame of nuclear matter at rest,
 where the $\eta$ self-energy is evaluated (see eq.(\ref{eqn:selfint})).
 We regularize the integral by means of a cut off rather than 
 by dimensional regularization as done in ref.\cite{inoue}.
 The choice of a cut off is preferable when one performs 
 the calculation in nuclear matter since Lorentz covariance is manifestly
 broken given the fact that one has a privileged frame of reference,
 the one where nuclear matter is at rest.
 The integration variable $q$ is the momentum of the meson in the loop,
 but in order to obtain a Lorentz invariant quantity 
 when $\rho = 0$ and $\vec P \ne 0$,
 the limits in eq.(\ref{eqn:labint}) must be implemented in such a way that
 the meson momentum in the rest frame of the original meson baryon state,
 $\vec q_{\mbox{\tiny cm}}$, should be smaller than the cut off taken.
 Another possibility is of course to make a boost and work in the
 meson baryon CM frame which is the option followed in ref.\cite{ramos}.
 Appropriate choices of the cut off and the subtraction constants($a_i$)
 give the equivalent loop functions as in dimensional regularization
 (see also ref.\cite{bennhold}).
 Evaluating the integral in CM frame of the meson baryon system,
 we find that the choice
\begin{equation}
   q_{\mbox{\tiny cm}}^{\mbox{\tiny max}} = 1 \mbox{GeV} ~,~~
   a_{\pi N}    =  34 \mbox{MeV} ,~~
   a_{\eta n}   =  14 \mbox{MeV} ,~~
   a_{K \Lambda}=  39 \mbox{MeV} ,~~
   a_{K \Sigma} = -22 \mbox{MeV}
\label{eqn:subconst}
\end{equation}
 gives meson baryon loop functions equivalent
 to those in our previous paper. 
 In order to find $G(P^0,\vec P; \rho)$ which 
 appears  in eq.(\ref{eqn:bseq}) we shall use the formulation
 of eq.(\ref{eqn:labint}) and eq.(\ref{eqn:subconst}) in the free case
 but will take into account Pauli blocking in the intermediate nucleon
 states, plus the meson and baryon self-energy in the intermediate states.

 The Pauli blocking is one of the important nuclear matter effects.
 We incorporate this by replacing the free nucleon propagator as
\begin{equation}
 \frac{1}{P^0 - q^0 - E_l(\vec P-\vec q)} 
 \to
 \frac{\theta(|\vec P-\vec q| - k_F)}{P^0 - q^0 - E_l(\vec P-\vec q)} 
\end{equation}
 with the unit step function $\theta(|\vec P-\vec q| - k_F)$,
 which prevents the scattering to intermediate nucleon states
 below the Fermi momentum.

 Another important nuclear matter effect is the dressing of hadrons.
 All the baryons and mesons in the intermediate loops interact 
 with nucleons of the Fermi sea and their dispersion relations are changed.
 For the baryons, we incorporate this effect,
 within a mean-field approach, as a momentum-independent binding potential.
 Due to baryon number conservation, only the difference
 between the nucleon and hyperon potentials is relevant for our purpose
 and is introduced in the hyperon propagator in the kaon-hyperon loops.
 For the mesons, we incorporate the dressing effect by introducing 
 the self-energy function $\Pi_l(q^0,\vec q ~; \rho)$.
 The meson propagators are replaced, using the Lehmann representation, as
\begin{eqnarray}
 \frac{1}{ (q^0)^2 - {\vec q\,}^2 - m_l^2}
 &\to&
 \frac{1}{ (q^0)^2 - {\vec q\,}^2 - m_l^2 - \Pi_l(q^0,\vec q ~; \rho)}
 \\
 &=&
 \int_0^{\infty} \! \! \! \! d \omega~
 2 \omega \frac{S_l(\omega,\vec q ~; \rho)}
               {(q^0)^2 - \omega^2 + i \epsilon}
\end{eqnarray}
  where $S_l(q^0,\vec q ~; \rho)$ is meson spectral function given by
\begin{equation}
  S_{l}(q^0, \vec q~; \rho)
  = 
  - \frac{1}{\pi} 
    \frac{\mbox{Im}[ \Pi_{l}( q^0,\vec q~;\rho) ]}
         { | (q^0)^2 - {\vec q\,}^2 - m_l^2 
                 -\Pi_{l}( q^0,\vec q~;\rho)|^2 } 
\end{equation}
 as a function of energy and momentum in the nuclear matter frame.

 Summarizing, the in-medium $\pi N$ loop function, for example,
 is calculated by 
\begin{eqnarray}
 & & G_{\pi N}(P^0, \vec P ~;\rho) 
 ~=~ a_{\pi N} + 
\\
& & \quad \quad \quad
\int \! \! \frac{d^4 q}{(2 \pi)^4} 
~\theta(q_{\mbox{\tiny cm}}^{\mbox{\tiny max}}-|\vec q_{\mbox{\tiny cm}}|)
\nonumber
\frac{M_N}{E_N(\vec P-\vec q)} 
\frac{\theta(|\vec P-\vec q| - k_F)}
     {P^0 - q^0 - E_N(\vec P-\vec q) + i\epsilon} 
\int_0^{\infty} \! \! \! \! \! \! d \omega
\frac{2 \omega S_{\pi}(\omega,\vec q~;\rho)}{(q^0)^2-\omega^2+i\epsilon}
\end{eqnarray}
 with 
\begin{equation}
 \vec q_{\mbox{\tiny cm}} = 
     \left[  \left( \frac{P^0}{\rts} - 1 \right) 
                 \frac{\vec P \cdot \vec q}{|\vec P|^2}
             - \frac{q^0}{\rts} \right] \vec P + \vec q
\end{equation}
 from the Lorentz boost.

 The 2-loop function needed for the $\pi \pi N$ channels is 
 also modified in nuclear matter. 
 We still take a zero constant for the real part according to 
 our previous study in the free space \cite{inoue}.
 On the other hand, the imaginary part is modified along
 the lines of $G_{\pi N}$ and we obtain
\begin{eqnarray}
   \mbox{Im}[ G_{\pi\pi N}(P^0, \vec P~;\rho) ]
   \!\!\! &=& \!\!\! 
   - 
   \int \!\! \frac{d^3 \vec q_1}{(2 \pi)^3} \! \!
   \int \!\! \frac{d^3 \vec q_2}{(2 \pi)^3} 
   (\vec q_1 - \vec q_2)^2 \frac{M_N}{E_N} 
   \theta\left( |\vec P-\vec q_1-\vec q_2|-k_F \right) 
   \\
   & & \quad \times
   \theta\left( P^0-E_N \right)
   \pi \! \int_0^{\infty} \! \! \! \! d \omega
   S_{\pi}(\omega, \vec q_1 ~; \rho)
   S_{\pi}(P^0-E_N-\omega, \vec q_2 ~; \rho)
   \nonumber
\end{eqnarray}  
  with $E_N \equiv E_N(\vec P - \vec q_1-\vec q_2)$
  when both the Pauli blocking and the dressed pion are considered.

\section{Input}

 We use commonly accepted values of baryon binding potentials
 which are $-70~\rho/\rho_0$ MeV for the nucleon 
 and $-30~\rho/\rho_0$ MeV for the hyperons.
 Hence, the difference $+40~\rho/\rho_0$ MeV is added
 to the hyperon energy $E_Y$ in the hyperon propagator.

 The effective $K$-$N$ interaction in the medium is studied
 in ref.\cite{wolfram,kaiser,angels} and the resulting kaon self-energy
 is $\Pi_{K}(\rho) = 0.13~ m_K^2~ {\rho}/{\rho_0}$ approximately.
 We use this real constant self-energy for kaons in the kaon-hyperon loops.

\begin{figure}[t]
  \centerline{ 
   \epsfysize = 40 mm  \epsfbox{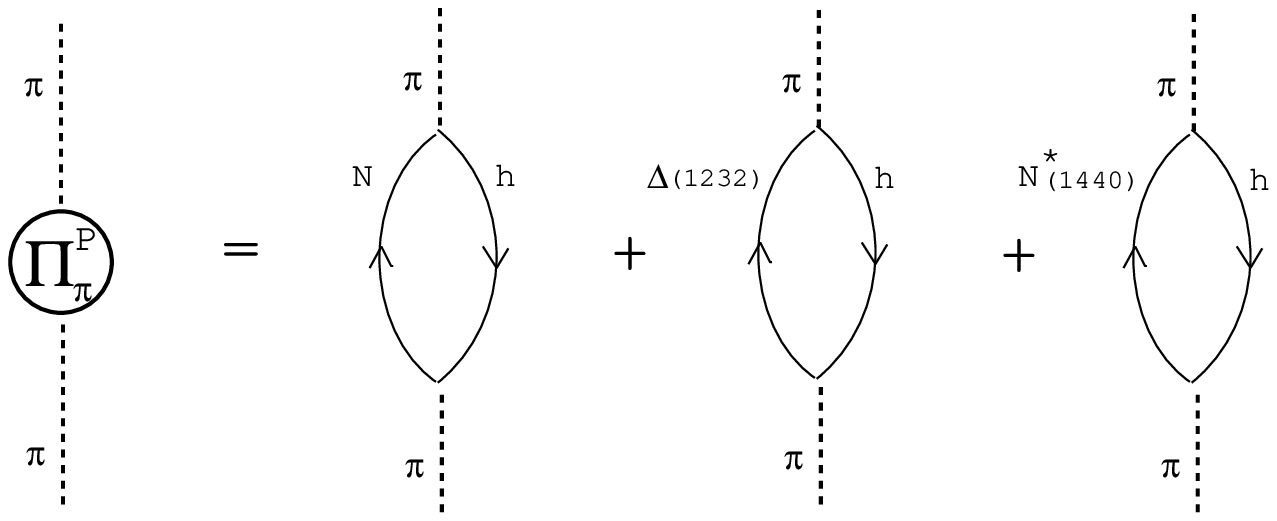}
   \hspace{20mm}
   \epsfysize = 40 mm  \epsfbox{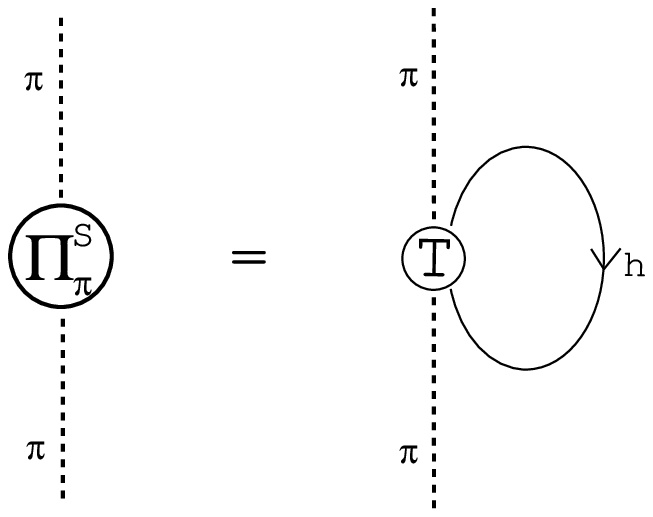}
              }
  \caption{
   Diagrammatic representation of the P-wave(left) and S-wave(right)
   term of pion self-energy. 
   The crossed diagrams are omitted in the figure.
          }
  \label{fig:piself}
\end{figure}

 The pion self-energy which we use in the $\pi N$ and $\pi\pi N$ loops, 
 consists of a P-wave term and an S-wave term as shown in Fig.\ref{fig:piself}.
 The P-wave term includes N-h, $\Delta(1232)$-h and $N^*(1440)$-h excitations, 
 and is basically given by
\begin{equation}
 \Pi_{\pi}^{P}(q^0, \vec q~;\rho) = 
 \left( \frac{f_{\pi NN}}{m_{\pi}} \right)^2  {\vec q\,}^2 
 \left\{          U_N(q^0,\vec q~;\rho) 
         + U_{\Delta}(q^0,\vec q~;\rho)
         +    U_{N^*}(q^0,\vec q~;\rho) \right\}
 \label{eqn:pipw}
\end{equation}
  where $f_{\pi \pi N} = 1.02$, and 
        $U_N(q^0,\vec q~;\rho)$, 
 $U_{\Delta}(q^0,\vec q~;\rho)$ and 
    $U_{N^*}(q^0,\vec q~;\rho)$ 
 are the Lindhard functions corresponding
 to the above excitations respectively.
 For example, $U_{N^*}(q^0,\vec q~;\rho)$ is given by
\begin{equation}
 U_{N^*}(q^0,\vec q~;\rho) =
 \frac32  \left( \frac{f_{\pi NN^*}}{f_{\pi N N}} \right)^2 \! \rho~
 \frac{M_{N^*}}{|\vec q| k_F}
 \left[  z +\frac12 (1-z^2)    \ln \frac{z+1}{z-1}
       + z'+\frac12 (1-{z'}^2) \ln \frac{z'+1}{z'-1}
 \right]
\label{eqn:unst}
\end{equation}
\begin{eqnarray}
 z  &=&  \frac{M_{N^*}}{|\vec q| k_F}
    \left(  q^0 - \frac{{\vec q\,}^2}{2 M_{N^*}} 
          -(M_{N^*} - M_N) + \frac{i}{2} \Gamma_{N^*}(q^0,\vec q\, ) 
    \right) 
\label{eqn:defz}
 \\
 z' &=&
   \frac{M_{N^*}}{|\vec q| k_F}
   \left( -q^0 - \frac{{\vec q\,}^2}{2 M_{N^*}}
          -(M_{N^*} - M_N) + \frac{i}{2} \Gamma_{N^*}(-q^0,\vec q\, ) 
   \right) 
\label{eqn:defzdash}
\end{eqnarray}
 where we use $f_{\pi NN^*}/f_{\pi N N} = 0.477$.
 This term is small compared to the $\Delta$-h excitation term
 whose coupling is $f_{\pi N \Delta}/f_{\pi N N} = 2.13$,
 but has a sizable contribution for energetic pions.
 The expression for $U_{\Delta}$ is analogous to that
 in eq.(\ref{eqn:unst}-\ref{eqn:defzdash}) and is given in ref.\cite{cordoba},
 the factor $3/2$ in eq.(\ref{eqn:unst}) is replaced by $2/3$ and 
 the coupling, mass and width of the $N^*$ are replaced 
 by the corresponding magnitudes for the $\Delta$.
 In the actual calculation
 we use the P-wave contribution eq.(\ref{eqn:pipw}) improving with
 a form factor, a short range correlation and a recoil factor as
\begin{eqnarray}
 & &
 \left( \frac{f_{\pi NN}}{m_{\pi}} \right)^2 {\vec q\,}^2 
 \sum_i U_i(q^0,\vec q~;\rho)
 \to
 \left( \frac{f_{\pi NN}}{m_{\pi}} \right)^2 {\vec q\,}^2 
 F(q^0,\vec q\,)^2
 \sum_i U_i(q^0,\vec q~;\rho)
 \\
 & &
 \sum_i U_i(q^0,\vec q~;\rho)
 \to
 \frac{ \sum_i U_i(q^0,\vec q~;\rho) }
      {1 - \left( \frac{f_{\pi NN}}{m_{\pi}} \right)^2  g'
           \sum_i U_i(q^0,\vec q~;\rho) }
 \\
 & &
 U_i(q^0,\vec q~;\rho)
 \to 
 \left\{
 \begin{array}{c}
 \left(1-\frac{q^0}{2 M_N}     \right)^2 U_N(q^0,\vec q~;\rho)
 \\
 \left(1-\frac{q^0}{M_{\Delta}}\right)^2 U_{\Delta}(q^0,\vec q~;\rho)
 \\
\left(1-\frac{q^0}{2 M_{N^*}} \right)^2 U_{N^*}(q^0,\vec q~;\rho)
 \end{array}
 \right.
\end{eqnarray}
 respectively, 
 where $F(q^0,\vec q\,)$ is a monopole form factor
 with cutoff of 1.2 GeV and $g'=0.6$.

 The S-wave term of the pion self-energy is taken in the 
 so called ``$T\rho$'' approximation,
\begin{equation}
   \Pi_{\pi}^{S}(q^0,\vec q~;\rho) 
 = T_{\pi N} \times  \rho 
 = \left\{ \frac13 T_{1/2} +  \frac23 T_{3/2} \right\} \times \rho 
\end{equation}
 using the free space S-wave $\pi$-$N$ scattering amplitudes
 calculated in the present model.
 This term includes the contribution of $N^*(1535)$-h excitation
 through the isospin 1/2 amplitude $T_{1/2}$,
 and is small compared to the P-wave one and
 becoming only sizable for very energetic pions
 with about 700 MeV momentum.

 \begin{figure}[t]
 \centerline{ 
 \epsfysize = 58 mm  \epsfbox{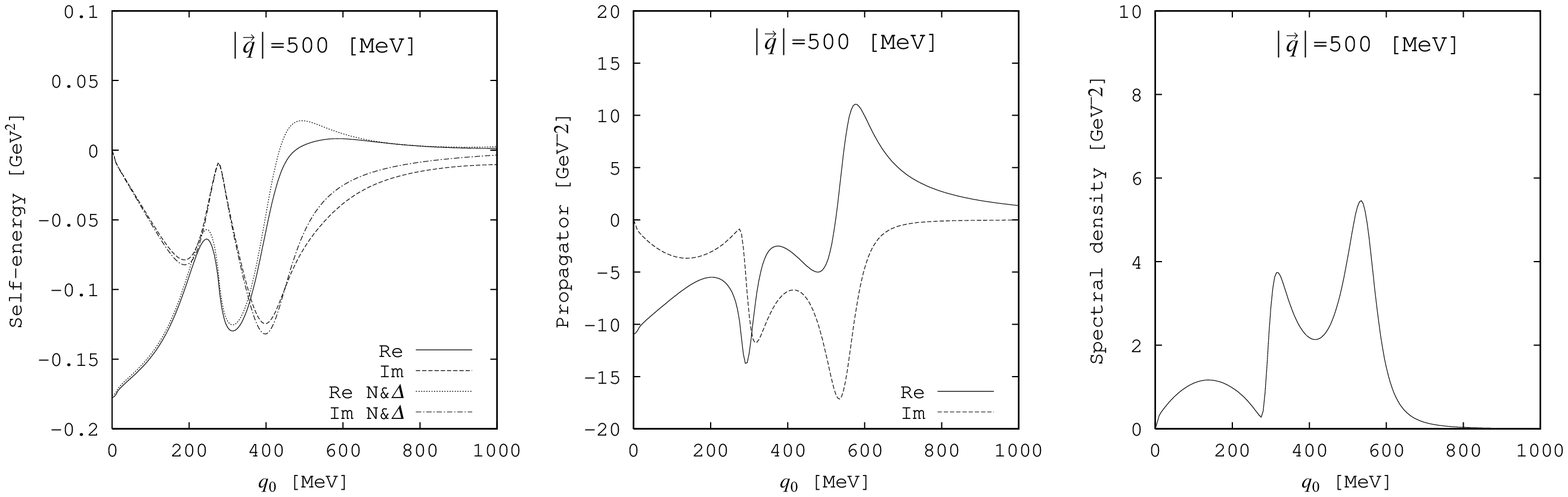}
              }
 \caption{
  Self-energy(left), propagator(center) and spectral function(right)
  of a pion for 500 MeV momentum, at normal nuclear matter density.
  The dotted and dashed dotted lines in the self-energy(left) correspond to the
  conventional ones including only the $N$-h and $\Delta$-h excitations.
         }
 \label{fig:piprop}
\end{figure}

 The total self-energy of pion is given by
 $\Pi_{\pi}^P(q^0,\vec q~;\rho) + \Pi_{\pi}^S(q^0,\vec q~;\rho)$
 and in Fig.\ref{fig:piprop} we show
 the self-energy(left), propagator(center) and spectral function(right)
 of the pion for 500 MeV momentum for normal nuclear matter density.
 In the self-energy(left) graph, the accompanying dotted and
 dashed dotted lines  correspond to the case when
 we take only the $N$-h and $\Delta$-h excitations.
 They show that the contribution of $N^*(1440)$ is relatively small.
 
 One of the needed input in our evaluation is the self-energy of the $\eta$
 which appears in the $\eta n$ loop. However, this is not known beforehand
 since this is what we want to calculate here. 
 We perform for this purpose a self-consistent calculation.
 In a first step, the $T\rho$ approximation is used since we have previously   
 calculated the free $\eta n$ T matrix. With this we obtain a new self-energy
 of the $\eta$ which is introduced in the second step,
 and so on until convergence is reached and one obtains
 an output $\eta$ self-energy equal to the input one.

\section{Results}

 \begin{figure}[t]
  \centerline{ 
  \epsfysize = 55 mm  \epsfbox{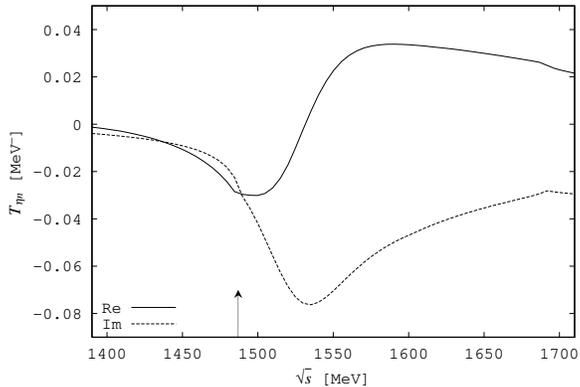}
              }
 \caption{
  The free space $T_{\eta n}$ amplitude as a function of
  the invariant energy $\rts$. The arrow shows the $\eta n$ threshold.
         }
 \label{fig:t66free}
\end{figure}

 Fig.\ref{fig:t66free} shows the free space $\eta$-neutron scattering
 amplitude $T_{\eta n}$ obtained in the present model. 
 It is plotted as a function of the invariant energy $\rts$,
 which is 1487 MeV at the threshold. 
 One sees that the amplitude changes drastically,
 even from attractive to repulsive, above the threshold.
 This is due to the coupling to the $N^*(1535)$ resonance
 which is generated dynamically in the present model.
 Note that the threshold is about 50 MeV below the center of the resonance.
 We can obtain the $\eta$ self-energy as an explicit function of $k^0$ and
 $\vec k$ as independent variables.
 One of the interesting applications is the study of $\eta$ bound states
 in the nucleus, where the $\eta$ will have small energy and momenta.
 For this purpose we can evaluate the $\eta$ self-energy 
 for $k^0=m_{\eta}$ and $\vec k=0$ as a good approximation.
 Non-local effects from the consideration of the energy and momentum
 dependence of the self-energy give
 very small corrections for weakly bound states 
 as demonstrated in ref.\cite{garcia2} for $K^-$ atoms.

\begin{figure}[p]
 \centerline{ 
 \epsfysize = 55 mm  \epsfbox{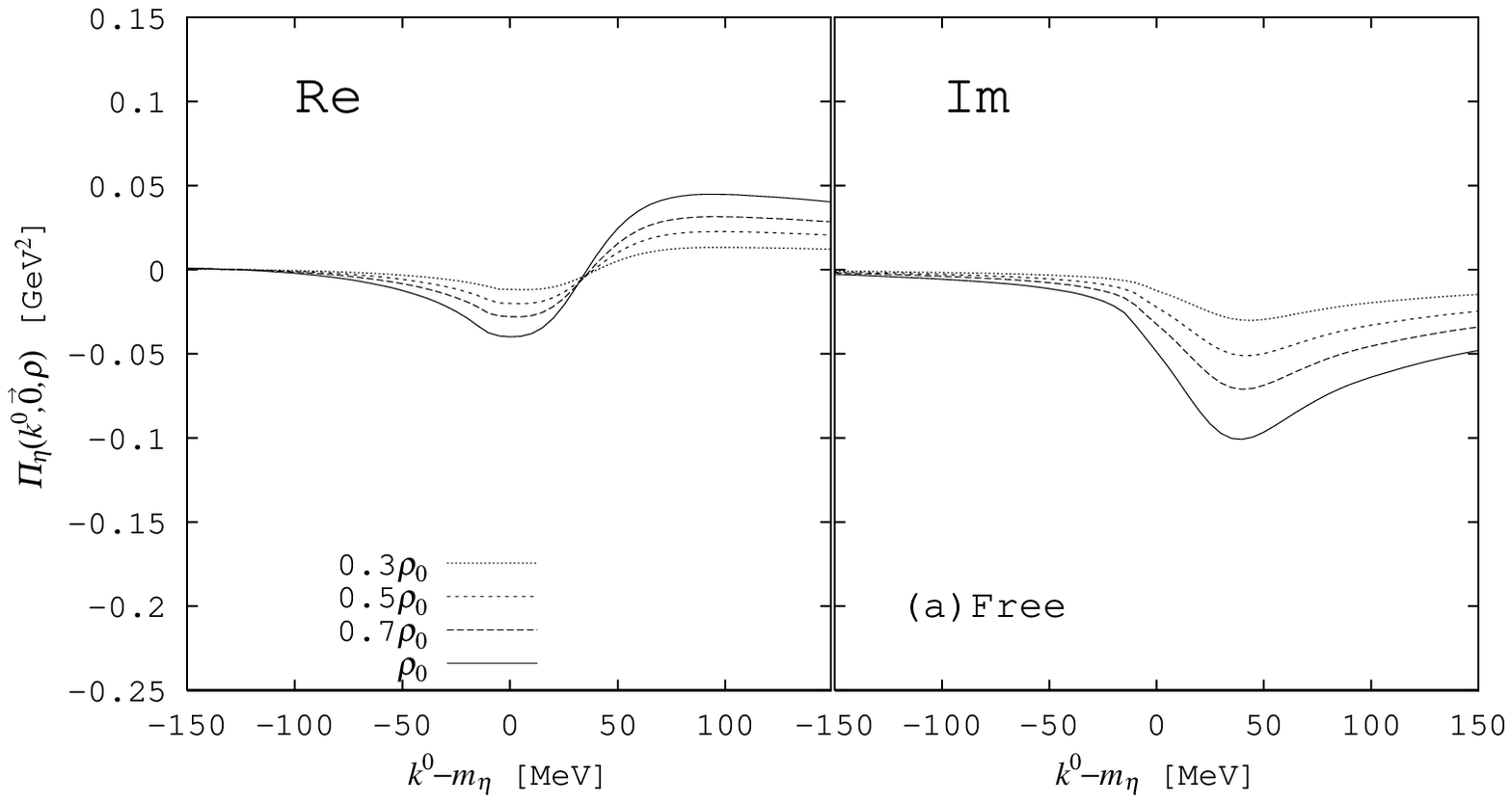}
              }
 \centerline{ 
 \epsfysize = 55 mm  \epsfbox{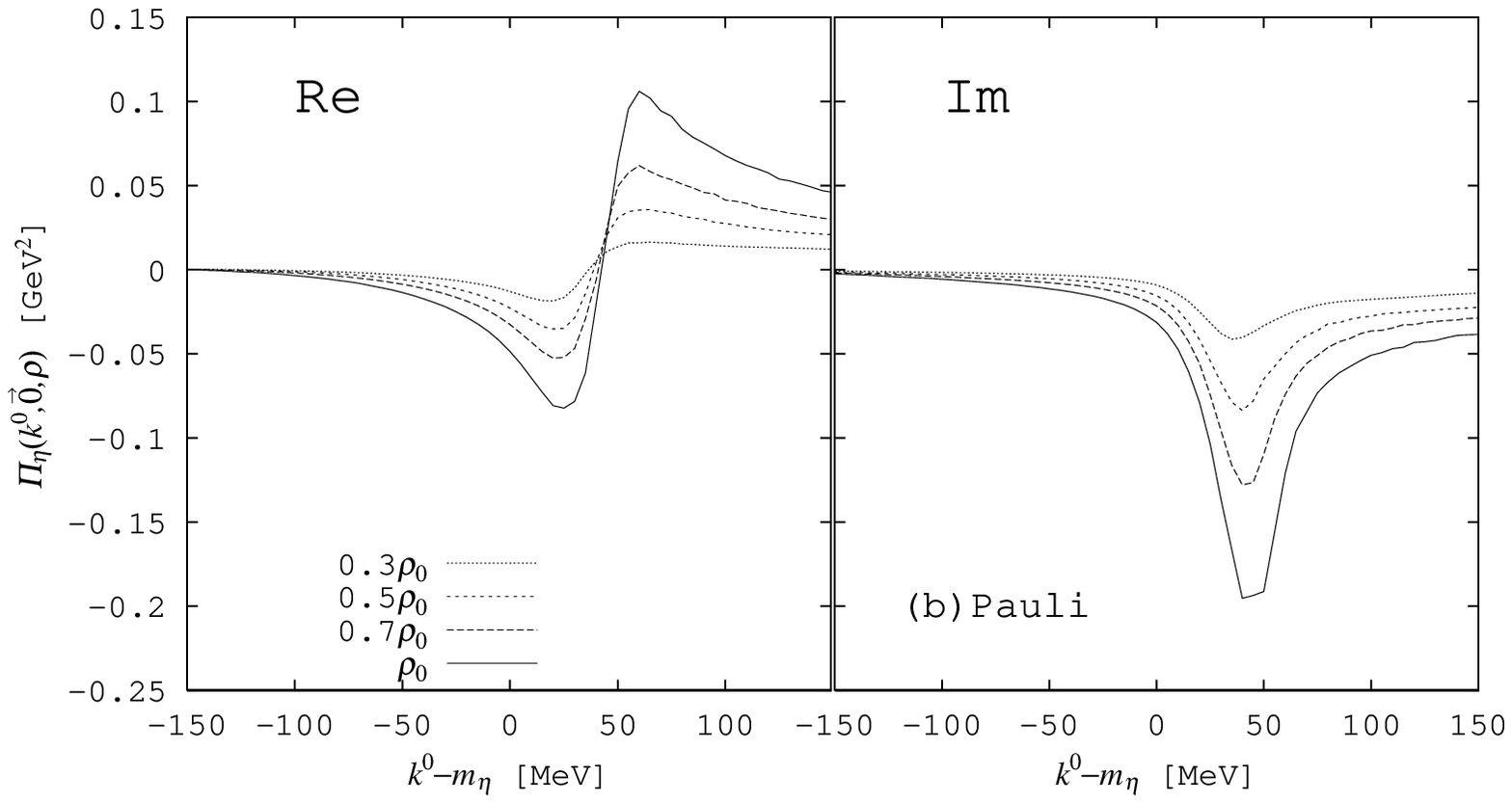}
              }
 \centerline{ 
 \epsfysize = 55 mm  \epsfbox{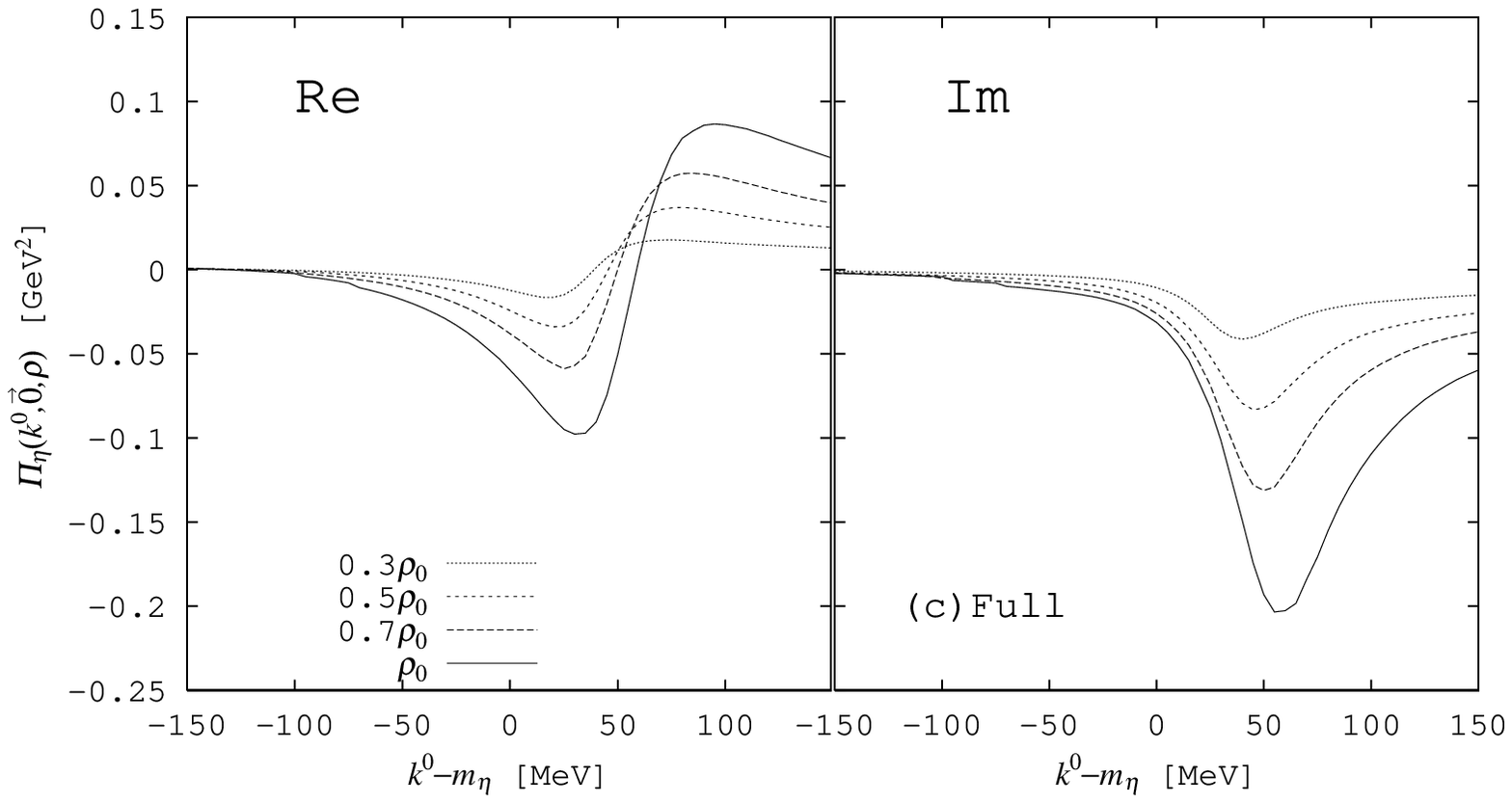}
              }
 \caption{
  $\eta$ self-energy for zero momentum as a function of energy,
  for four different densities and in three different approximations.
  Top, (a) Free:  the free space $\eta$-$N$ amplitude is used.
  Middle, (b) Pauli: the Pauli blocking is taken into account.
  Bottom, (c) Full:  both the Pauli blocking and hadron dressing 
  are taken into account.
         }
 \label{fig:iiabc}
\end{figure}

 In order to evaluate the $\eta$ self-energy in a first step,
 we substitute the free $\eta n$ amplitude
 in eq.(\ref{eqn:selfint}), setting $\vec k = \vec 0$, and obtain
 the $\eta$ self-energy for zero momentum shown in  Fig.\ref{fig:iiabc} top. 
 This simplest calculation is nearly equal to the $T\rho$ approximation.
 The magnitude of the self-energy increases almost linearly in $\rho$
 and its shape reflects that of $T_{\eta n}$.   

 Then we turn on the Pauli blocking in the way explained in section 2.
 The resulting $\eta$ self-energy for zero momentum is shown in
 Fig.\ref{fig:iiabc} middle. Comparing to the top one, 
 we see the interesting effects of the Pauli blocking.
 First of all, the resonant shape is strongly enhanced.
 This can be understood as the reduction of the decay width of the resonance,
 since the Pauli blocking forbids the decay to nucleons
 with momentum smaller than $k_F$.
 Second, the peak of the resonance, 
 seen in the imaginary part of the amplitudes,
 or equivalently in the zero of the real part, is only shifted
 moderately to higher energies.
 This is in contrast with the $\bar K$ case,
 where the $\Lambda(1405)$ resonance in the $\bar K N$ interaction,
 is about 80 MeV shifted upward \cite{waas,ramos}.
 As pointed out in ref.\cite{waas}, this is natural because 
 a large part of the $N^*(1535)$ resonance generated in the present model,
 comes from the $K^+ \Sigma$, $K^0 \Sigma^0$ and $K^0 \Lambda$ components
 \cite{siegel,kaiser,inoue}, which have nothing to do with Pauli blocking. 
 In ref.\cite{waas} it is argued that the Fermi motion compensates 
 the decay width reduction due to the Pauli blocking.
 There also is some effect of this sort here, 
 but we still find a net reduction of the $N^*(1535)$ width due to the
 consideration of both effects, resulting in an enhanced strength of 
 both the real and imaginary parts of the $\eta$ self-energy.
 A major difference in our model from ref.\cite{waas}, 
 at the present stage of the calculation,
 is the presence of the 3-body $\pi\pi N$ channels.
 Also, our free space $T_{\eta n}(\rts)$ shown in Fig.\ref{fig:t66free},
 is already different from that in ref.\cite{waas},
 where the peak of the resonance is seen at around 1590 MeV.

\begin{figure}[t]
 \centerline{ 
 \epsfysize = 60 mm  \epsfbox{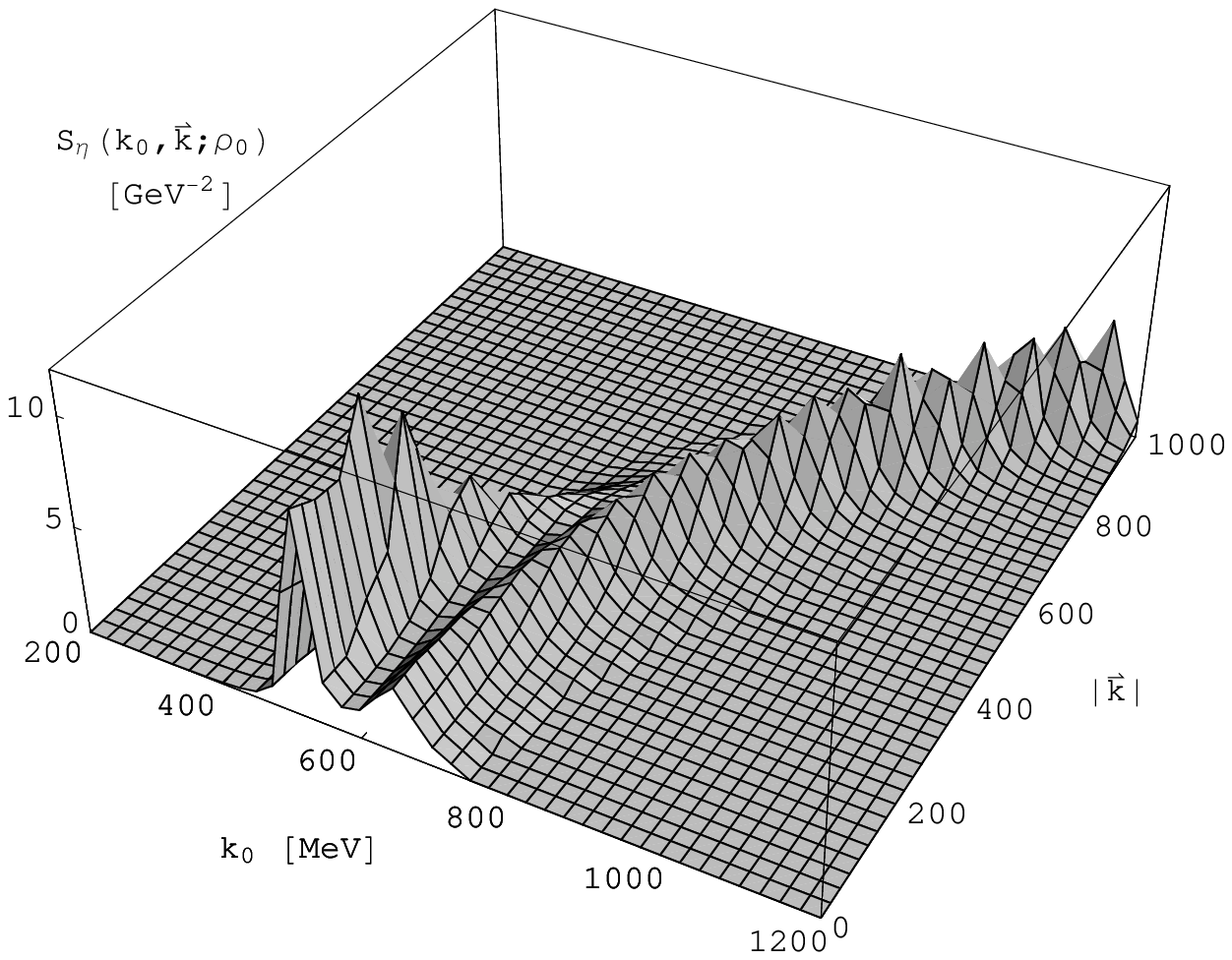}
 ~~~
 \epsfysize = 55 mm  \epsfbox{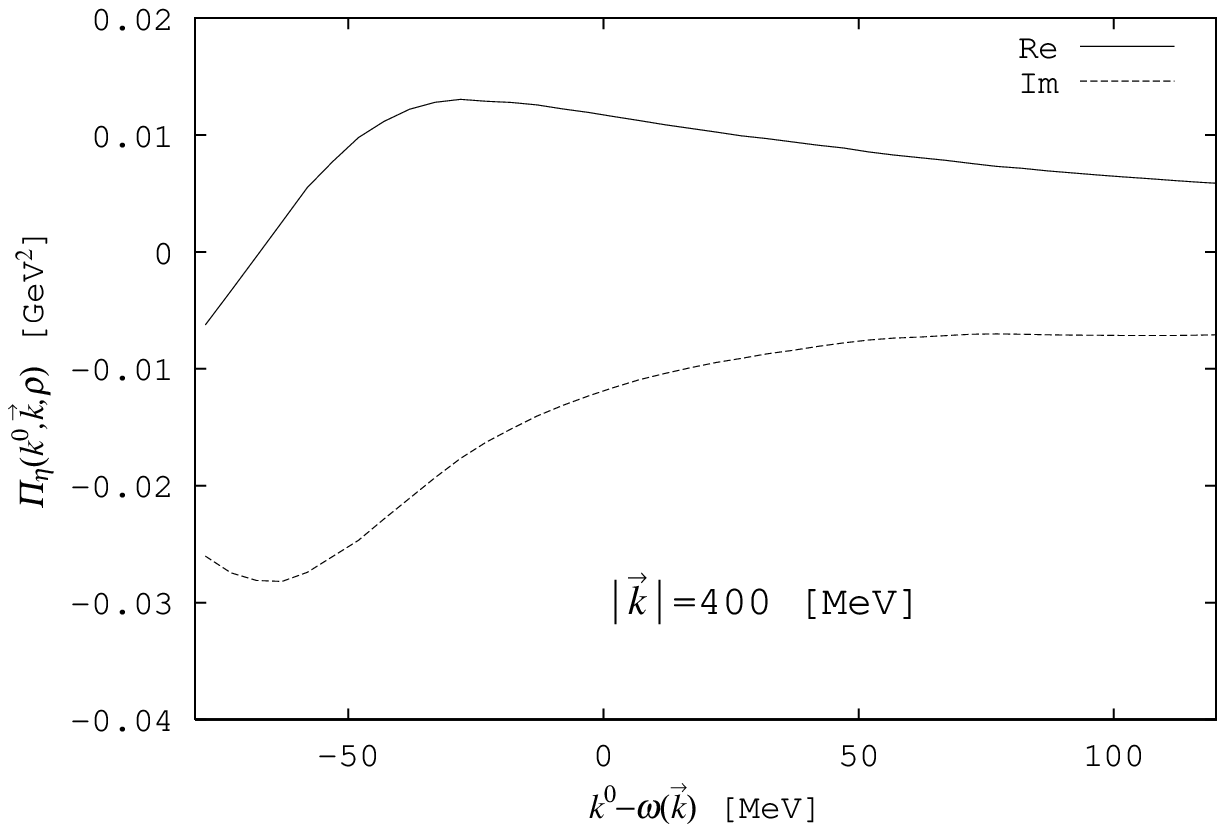}
              }
 \caption{
   Left: A rough sketch of the $\eta$ spectral function
   at normal nuclear matter density,
   where the peak value is not realistic because of the plotting resolution.
   Right: Self-energy of the $\eta$ with 400 MeV momentum
   at normal nuclear matter density as a function of energy.
         }
 \label{fig:sp2d}
\end{figure}

 Finally, we turn on the dressing of the hadrons
 in addition to Pauli blocking.
 This study is the main novelty of the present work.
 We use the input explained in the previous section.
 Besides this, we need the spectral function of the $\eta$ 
 in order to calculate the $\eta n$ loop function.
 As stated in the previous section,  we determine it in
 a self-consistent way by iterating the Bethe-Salpeter equation. 
 This means that we need to study the finite momentum case
 even if we are interested only in the $\eta$ with zero momentum.
 We, hence, evaluate the $\eta$ self-energy also for finite momentum
 and, for example, Fig.\ref{fig:sp2d} left shows a rough sketch of 
 the $\eta$ spectral function obtained at normal nuclear matter density
 as a function of $k^0$ and $\vec k$.
 The results in what follows include the dressed $\eta$ 
 accounting for the $\eta$ self-energy as a function $k^0$ and $\vec k$.

 The $\eta$ self-energy for zero momentum obtained 
 in this self-consistent approach is shown in Fig.\ref{fig:iiabc} bottom.
 One can see the effects of the hadron dressing in comparison to
 the middle figure of the panel.
 Apparently, the strength is spread wider.
 This is expected because the $\eta$ spectrum is distributed as stated above.
 Even then, the resonant shape is still clearly visible.
 The center of the resonance is shifted upward,
 for example, about 25 MeV for normal nuclear matter density.
 This means that the sum of the repulsion on $K$ and so on,
 surpasses the attraction on the pions.
 The difference of the hyperon binding energy and the nucleon one, 
 works as a repulsion for this calculation.
 The dressed $\eta$ also shifts it upward slightly 
 because the $\eta$ with large momentum(larger than about 250 MeV)
 feels a weak repulsion, as shown in Fig.\ref{fig:sp2d} right.
 Recall that the $K \Sigma$, $K \Lambda$ and $\eta n$ components are 
 more important than the $\pi N$ and $\pi \pi N$ components 
 for the resonance in the present model.
 The 25 MeV shift for normal nuclear matter density
 is small(less than 2\% of the $N^*(1535)$ mass)
 and consistent with $\eta$ photo-production data \cite{landau},
 which together with the theoretical analysis of ref.\cite{carrasco}
 assuming no shift of the resonance, suggest that the in-medium mass
 of the resonance is almost the same as in free space.
 
\begin{figure}[t]
 \centerline{ 
 \epsfysize = 55 mm  \epsfbox{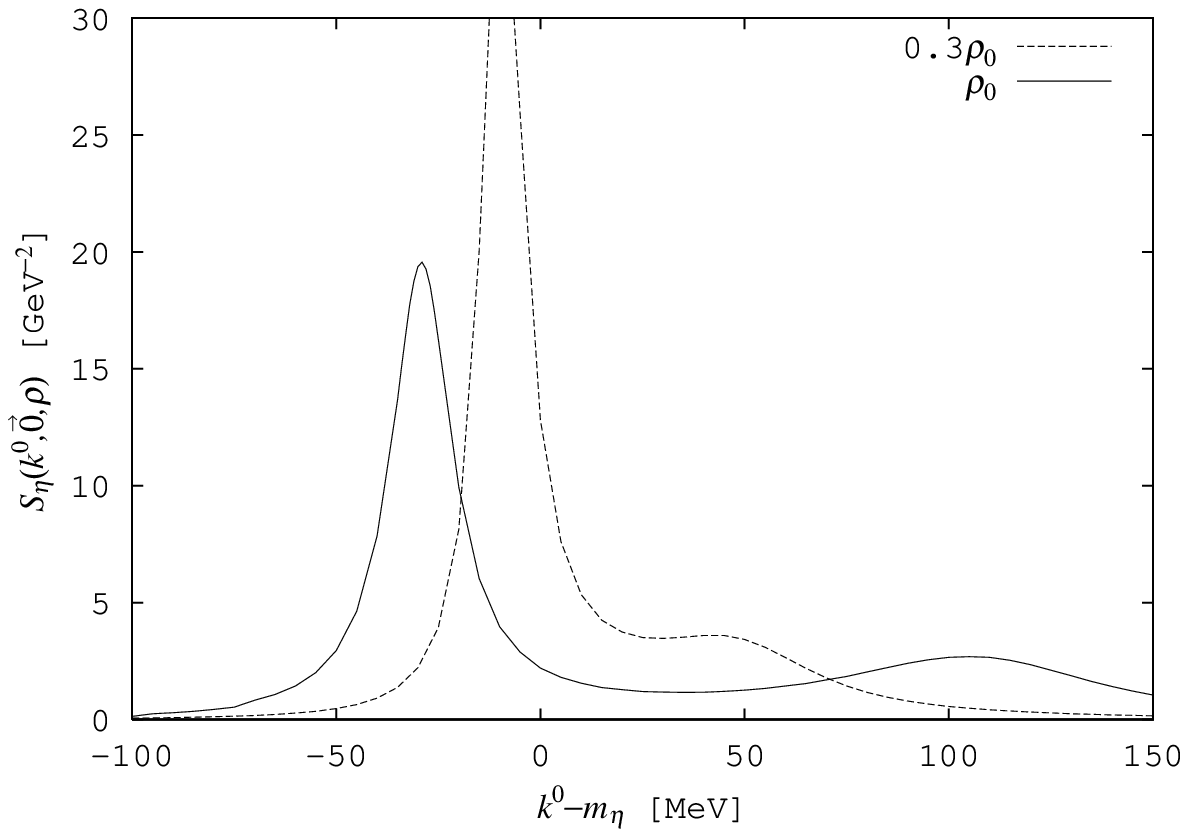}~~
 \epsfysize = 55 mm  \epsfbox{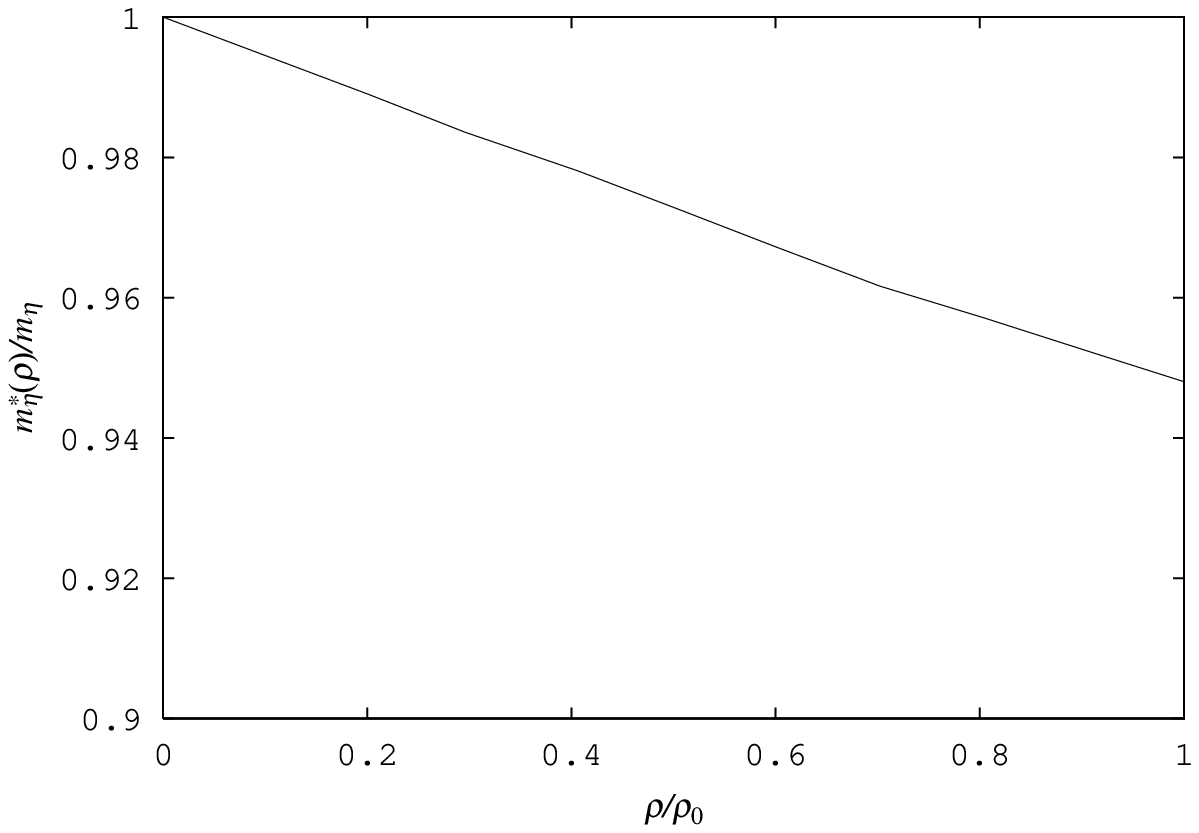}
              }
 \caption{ 
    Left:  $\eta$ spectral density for the zero momentum.
    Right: Effective mass of the $\eta$ as a function of density.
         }
 \label{fig:spfull}
\end{figure}

 Fig.\ref{fig:spfull} left shows the spectral function
 of a zero momentum $\eta$ corresponding to the final self-energy.
 We see still a narrow peak even for normal nuclear matter density and,
 hence, it makes sense to regard the $\eta$ in the medium
 as a quasi-particle with a modified mass.
 It is interesting to note a second peak in the spectral function 
 at higher energies, which, as noted in ref.\cite{waas}, comes from
 the coupling of the $\eta$ to the $N^*(1535)$-h.
 We can observe in the figure that as the density increases 
 the two peaks move away from each other.

 We define the effective mass $m_{\eta}^*(\rho)$ as
 the energy $k^0$ which satisfies
\begin{equation}
   (k^0)^2 - m_{\eta}^2 - \mbox{Re}[\Pi_{\eta}(k^0,\vec{0}~;\rho)]=0 ~,
\label{eqn:effmass}
\end{equation}
 which is almost the position of the peak in the spectral function.
 The result is plotted in Fig.\ref{fig:spfull} right,
 which indicates that approximately one has
\begin{equation}
   \frac{m_{\eta}^*}{m_{\eta}} \simeq 1 - 0.05 \frac{\rho}{\rho_0}
\end{equation}   
 in short. This moderate downward shift agree with the result of \cite{waas}.
 While, a much stronger, about 11\%, downward shift
 for normal nuclear matter density, is reported in \cite{tsushima} where  
 the quark meson coupling model is used.
 It should be interesting to test these predictions from experimental
 data of $\eta$ mesic nuclei.

\begin{figure}[t]
 \centerline{ 
 \epsfysize = 55 mm  \epsfbox{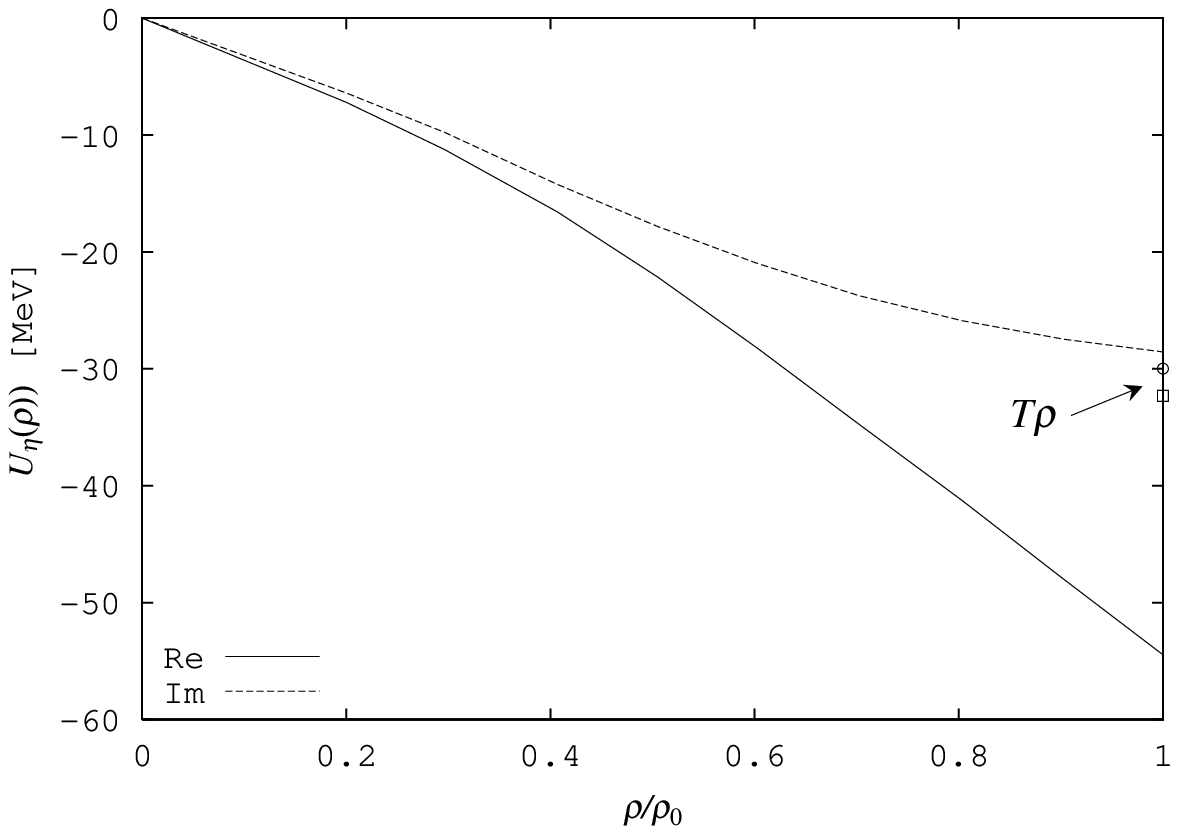}~~
 \epsfysize = 55 mm  \epsfbox{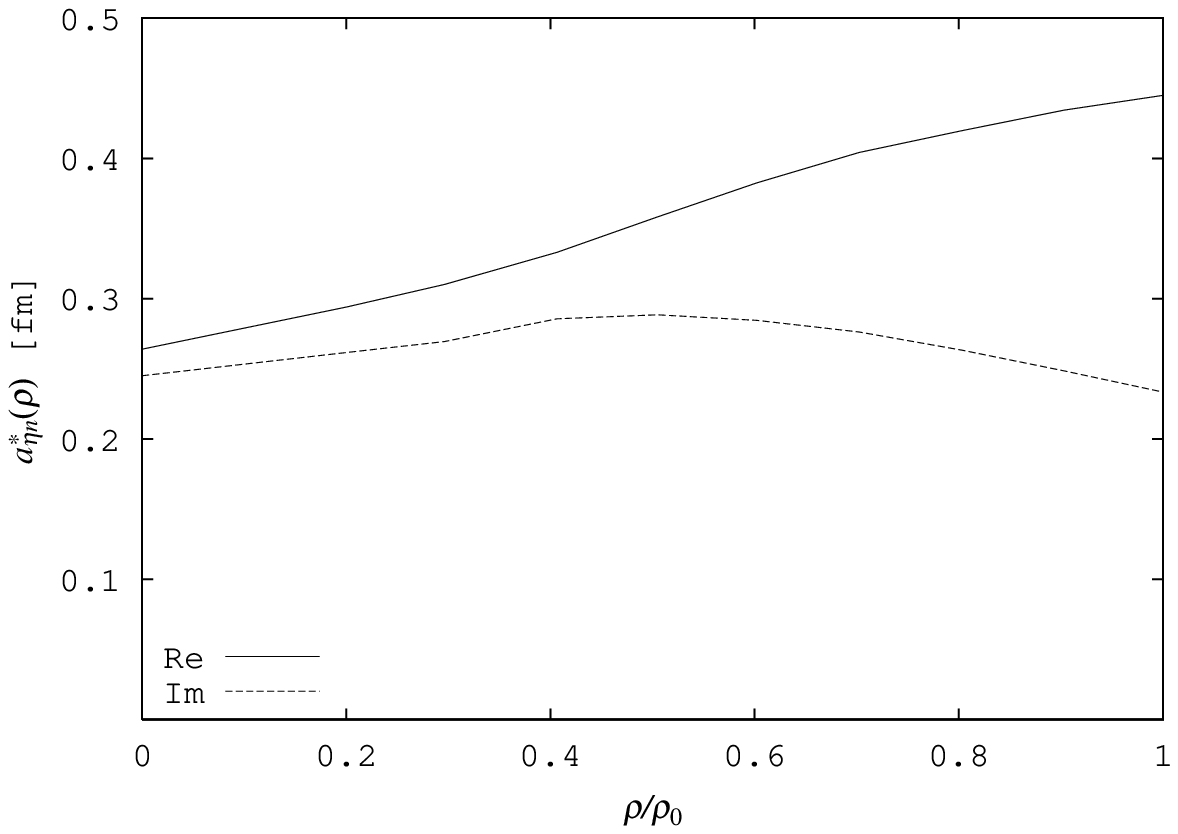}
             }
 \caption{
   Left: $\eta$ optical potential as a function of density.
   The box and circle stand for the real and imaginary parts
   of the potential in the $T\rho$ approximation 
   with the threshold $T_{\eta n}$ corresponding to 
   a scattering length of $0.26 + i0.24$ fm \cite{inoue}.
   Right: $\eta$-$n$ effective scattering length as a function of density.
         }
 \label{fig:optfull}
\end{figure}

 The optical potential $U_{\eta}(\rho)$ and 
 the effective $\eta$-neutron scattering length $a_{\eta n}^*(\rho)$ defined as
\begin{eqnarray}
  U_{\eta}(\rho) &=& \frac{ \Pi_{\eta}(m_{\eta},\vec{0},\rho) }{2  m_{\eta}}
  \\
  a_{\eta n}^*(\rho) &=& -\frac{1}{4 \pi}\frac{M_n}{M_n + m_{\eta}}
                    \frac{\Pi_{\eta}(m_{\eta},\vec{0},\rho) }{\rho}
\end{eqnarray}
 are useful when we study the bound state of an $\eta$ in nuclei
 using the local density approximation.
 We plot them in Fig.\ref{fig:optfull}.
 One can see that the real and imaginary parts 
 of the optical potential deviate as the density increases.
 The real part becomes larger than the linear $T \rho$ approximation 
 while the imaginary part is nearly $T \rho$.
 This behavior is better seen in the effective scattering length,
 where the real part increases and the imaginary part almost stays the same. 
 The depth of the optical potential at normal nuclear matter density
 is $-54 - i 29$ MeV. This value is not in contradiction
 with the 5\% (28 MeV) mass shift of the $\eta$,
 because the self-energy is evaluated at an $\eta$ energy equal
 to the $\eta$ mass, 
 while the effective $\eta$ mass, $k^0$ in eq.(\ref{eqn:effmass}),
 is obtained using the $\eta$ self-energy calculated at same $k^0$ energy.
 It is interesting to compare these result with the ones obtained 
 in the literature.
 In ref.\cite{waas} the potential obtained was
 $U_{\eta}(\rho) \simeq (-20 - i 22) ~\rho/\rho_0 ~\mbox{MeV}$
 which provides an imaginary part similar to ours
 but the real part is about one half of the one obtained here.
 In ref.\cite{chiang} the potential obtained assuming that 
 the mass of the $N^*(1535)$ did not change in the medium,
 as we showed it was approximately the case here, was
 $U_{\eta}(\rho) = (-34 - i 24) ~\mbox{MeV}$ at $\rho=\rho_0$,
 somewhere in between the two former results.

 In order to facilitate the task of making an accurate as possible prediction 
 for the $\eta$ bound states in nuclei,
 we parameterize our results for the $\eta$ self-energy 
 as a function of energy and density.
 Given the strong energy dependence of the $\Pi_{\eta}(k^0,\vec 0~;\rho)$ 
 seen in Fig.\ref{fig:iiabc} for $k^0 - m_{\eta} < 0$,
 the consideration of this explicit energy dependence  in the Klein-Gordon
 equation should be important.
 We can parameterize our results in the
 region $-50 ~\mbox{MeV} < k^0 - m_{\eta} < 0$, as
\begin{eqnarray}
 \mbox{Re}[ \Pi_{\eta}(k^0, \vec 0~; \rho) ] &=&   
       a(\rho) + b(\rho)(k^0-m_{\eta}) 
     + c(\rho)(k^0-m_{\eta})^2+ d(\rho)(k^0-m_{\eta})^3
\\
 \mbox{Im}[ \Pi_{\eta}(k^0, \vec 0~; \rho) ] &=&   
       e(\rho) + f(\rho)(k^0-m_{\eta}) 
     + g(\rho)(k^0-m_{\eta})^2+ h(\rho)(k^0-m_{\eta})^3
\end{eqnarray}
with 
\begin{eqnarray}
 a(\rho)&=& -36200.3   ~\rho/\rho_0 -24166.6   ~\rho^2/\rho_0^2 ~\mbox{MeV}^2
 \\
 b(\rho)&=& -1060.05   ~\rho/\rho_0 -326.803   ~\rho^2/\rho_0^2 ~\mbox{MeV}
 \\
 c(\rho)&=& -13.2403   ~\rho/\rho_0 -0.154177  ~\rho^2/\rho_0^2
 \\
 d(\rho)&=& -0.0701901 ~\rho/\rho_0 +0.0173533 ~\rho^2/\rho_0^2 ~\mbox{MeV}^{-1}
 \\
 e(\rho)&=& -43620.9   ~\rho/\rho_0 +11408.4   ~\rho^2/\rho_0^2 ~\mbox{MeV}^{2}
 \\
 f(\rho)&=& -1441.14   ~\rho/\rho_0 +511.247   ~\rho^2/\rho_0^2 ~\mbox{MeV}
 \\
 g(\rho)&=& -27.6865   ~\rho/\rho_0 +10.0433   ~\rho^2/\rho_0^2 
 \\
 h(\rho)&=& -0.221282  ~\rho/\rho_0 +0.0840531 ~\rho^2/\rho_0^2 ~\mbox{MeV}^{-1}
 ~~~.
\end{eqnarray}
 By using the Klein-Gordon equation, and substituting $\rho \to \rho(r)$
 in the spirit of the local density approximation, one can then obtain 
 binding energies and widths of the $\eta$ state in different nuclei.

\section{Conclusion}

 In the present paper we have used a chiral unitary approach
 successfully applied to the study of the $\pi N$ interaction 
 and its coupled channels, in particular the $\eta N$ channel,
 in order to evaluate the $\eta$ self-energy in a nuclear medium.
 We have taken into account the standard many body effects like
 Fermi motion and the Pauli blocking in the intermediate $N$ states. 
 In addition we have also included the self-energy of the baryons
 and mesons in the intermediate states, including the $\eta$ self-energy
 in a self-consistent way.
 The results obtained are interesting.
 While qualitatively similar to other ones obtained before, 
 we obtain however a deeper potential with a real part about
 twice as big as in former, more simplified, studies.
 In addition we show that the energy dependence of the $\eta$ self-energy
 is very pronounced below the $\eta$ threshold and
 it would be interesting to consider it in studies of $\eta$ bound
 states in nuclei. For this purpose we have parameterized our results
 in an easy analytical form which can be used 
 to solve the Klein-Gordon equation for $\eta$ bound states
 or to interpret some physical processes 
 where $\eta$ states close to threshold play some role \cite{khemchandani}.
 The stronger real part of our potential compared to previous ones
 and the fact that the imaginary part of the potential decreases rapidly
 as the $\eta$ energy decreases, open new hopes that $\eta$ mesic states, 
 relatively wide, but narrow enough compared to the binding energy, 
 could exist and be identified in actual experiments.
  
\section*{Acknowledgments}
 We would like to thank Prof. M.J. Vicente Vacas for useful discussions.
 This work has been partly supported by the Spanish Ministry of Education
 in the program 
 ``Estancias de Doctores y Tecn\'ologos Extranjeros en Espa\~na'',
 by the DGICYT contract number BFM2000-1326
 and by the EU TMR network Eurodaphne, contact no. ERBFMRX-CT98-0169.

\end{document}